\documentclass[10pt]{iopart}
\usepackage{iopams}

\expandafter\let\csname equation*\endcsname\relax
\expandafter\let\csname endequation*\endcsname\relax

\usepackage{amssymb}
\usepackage{amsmath}
\usepackage{soul}
\usepackage{cite}
\usepackage{enumerate}
\usepackage{empheq}
\usepackage{fancybox}
\usepackage{etoolbox}
\usepackage{hyperref}
\usepackage{color}
\makeatletter
\def\@mkboth#1#2{}
\newlength\appendixwidth
\preto\appendix{\addtocontents{toc}{\protect\patchl@section}}
\newcommand{\patchl@section}{%
  \settowidth{\appendixwidth}{\textbf{Appendix }}%
  \addtolength{\appendixwidth}{1.5em}%
  \patchcmd{\l@section}{1.5em}{\appendixwidth}{}{\ddt}%
}
\makeatother

\begin{document}

\title[]{Time-dependence of the effective temperatures of a two-dimensional Brownian gyrator with cold and hot components}

\author{Sara Cerasoli$^{1}$,Victor Dotsenko$^{2}$, Gleb Oshanin$^{2}$ \& Lamberto Rondoni$^{3,4}$}  
\address{$^1$ Department of Civil and Environmental Engineering, Princeton University, 54 Olden Street,
Princeton, NJ 08544, USA.}
\address{$^2$ Sorbonne Universit\'e, CNRS, Laboratoire de Physique Th\'eorique de la Mati\`{e}re Condens\'ee, UMR CNRS 7600,
75252 Paris Cedex 05, France}
\address{$^3$ Dipartimento di Scienze Matematiche, Dipartimento di Eccellenza 2018-2022, Politecnico di Torino, Corso Duca degli Abruzzi 24, 10129 Torino, Italy}
\address{$^4$ INFN, Sezione di Torino, Via P. Giuria 1, 10125 Torino, Italy}
  
\date{\today}

\begin{abstract}
We consider a model of a two-dimensional molecular machine - called Brownian gyrator -- that consists of two coordinates coupled to each other and  to separate 
heat baths at temperatures respectively $T_x$ and $T_y$. We consider the limit in which one component is passive, because its bath is ``cold'', $T_x \to 0$,
while the second is in contact with a ``hot'' bath, $T_y > 0$, hence it entrains the passive component in a stochastic motion. 
We derive an asymmetry relation as a function of time, from which time dependent effective temperatures can be obtained for both components. 
We find that the effective temperature of the passive element tends to a constant value, which is a fraction of  $T_y$, while the effective temperature 
of the driving component grows without bounds, in fact exponentially in time, as the steady-state is approached. 
 \end{abstract}


\maketitle

\section{Introduction}

The study of nonequilibrium deterministic and stochastic systems, along with their transient as well as steady-states properties, has gained 
momentum in the last few decades, both from the theoretical and the experimental point of view \cite{Kubo91,EvMo,Gall14,Ciliberto}.
Particular attention has been devoted to nonequilibrium fluctuations \cite{ECM2,GGBol,Jarz,ES2002,J2004,Bettolo} and to the violations 
of the Fluctuation Dissipation Relation and the notion of effective temperatures \cite{Kirkp,Dorfm,Ortiz,Cugli97,ContiD}.
A framework within which these developments are widely applied and investigated is that of artificial or natural molecular machines, 
which includes the so-called active particles, see {\em e.g.}\ Refs.\cite{peter,prost,Scholz}). Further avenues of investigation 
include fundamental notions such as the non-reciprocal non-Newtonian interactions emerging in out-of-equilibrium systems, cf.\ 
\cite{nos1,nos2,carl1,vinc,carl2,oli} and references therein. 
In some cases explicit relations have been rigorously derived within appropriate mathematical formalisms;  
this is the case of various kinds of fluctuation relations, of thermodynamic inequalities, and also of various 
kinds of fluctuation-dissipation relations in the linear response regime and beyond, 
see {\em e.g.}\  Refs. \cite{evans,Lam1,Lam2,Bettolo,Seki,Udo,pug,ol1,ol0,ol,neri,neri2}. 
\\
In this endeavor, the analysis of simple examples or {\em toy models} has been essential, in order to understand the properties
of realistic physical systems. Toy models are indeed amenable to exact analysis, and have often revealed unexpected behaviours, 
which would have been hard to discover from general relations or in fully realistic systems.  
\\
One such model is known as Brownian gyrator  (BG) -- a minimal molecular machine with stochastic dynamics which produces (on average)
a rotational motion in nonequilibrium conditions.  As detailed in Sect.\ref{model}, the BG consists of two time dependent components,
whose state, or position at time $t$, is given by the variables $(x_t,y_t)$. These components of the BG are
subject to thermal noises, characterised by the temperatures $T_x$ and $T_y$, respectively, and are coupled by a parabolic potential.
Mathematically, the BG is a pair of coupled Ornstein-Uhlenbeck processes.
Introduced in the physical literature by Ref. \cite{pel}, this model was later taken as a model of a nanoscopic heat engine \cite{1}. Indeed, for 
$T_x \neq T_y$, a non-zero (on average) torque is steadily produced, making the BG two-dimensional trajectory $\rho_t =(x_t,y_t)$ perform 
(on average) a rotational motion around the origin.
This interpretation of the BG has attracted a great deal of interest, motivating both experimental and analytical investigations, cf.\ {\em e.g.}\
Refs.\cite{al1,al2,crisanti,2,12,11,david,3,Lam3,fakhri1,fakhri2,komura,sesta,nas}. Extensions with more complicated 
non-linear couplings have also been considered \cite{kamenev,njp}.
\\
In this paper, we develop the analysis of Ref. \cite{Lam3}, in which each component of the BG model is subject to a regular external \textit{bias}, 
in addition to their usual Gaussian zero mean white-noises. The response of the system to such a bias was then analysed by an asymmetry 
function, analogous to the argument of the fluctuation relation, {\em i.e.}\ the logarithm of the ratio of the probabilities of opposite positions in the
plane $(x,y)$. This quantity was evaluated in the steady-state, thus identifying  the effective temperatures of both components and their limiting
values. The rather surprising result was that when one of the temperatures is set to zero, hence the corresponding component is passive
and entrained in motion by the second component, the effective temperature of the latter appears to be infinitely large. We are going to focus 
on this peculiar case, analyzing the temporal evolution of both effective temperatures.
\\
The paper is organized as follows: In Sec. \ref{model}, we define our biased BG and the properties to be studied; we
also show that the model at hand is identical, upon some appropriate rescaling of the parameters, to the one comprising 
two harmonically coupled beads, kept at two different temperatures and tethered to two walls \cite{al1,al2,fakhri1,fakhri2,komura}.  
In Sec. \ref{results}, we present the details of our derivations, and we illustrate our results. Finally, in Sec. \ref{conc} we summarize and 
comment on our findings.

\section{Model}
\label{model}
The \textit{biased} BG we consider here is described by two coupled Langevin equations in presence of a constant external force:
\begin{align}
\label{1}
\frac{dx}{dt} = F_x  - \frac{\partial U_0(x,y)}{\partial x} + \zeta_x(t) \,, \nonumber\\
\frac{dy}{dt} = F_y  - \frac{\partial U_0(x,y)}{\partial y}  + \zeta_y(t) \,,
\end{align}
where $F_x$ and $F_y$ are the components of the bias ${\bf F} = (F_x,F_y)$, $U_0$ expresses the coupling between the two components
and a confining parabolic potential:
\begin{align}
\label{pot}
U_0(x,y) =  \frac{x^2}{2} + \frac{y^2}{2} + u x y  \,,
\end{align}
where $0 < |u| < 1$ is a fixed parameter. Moreover,
$ \zeta_x(t)$ and $ \zeta_y(t)$ are statistically-independent Gaussian white-noises with zero mean and the covariance functions given by
\begin{align}
& \overline{\zeta_x(t_1) \zeta_x(t_2)} = 2 T_x \delta(t_1 - t_2) \,, \nonumber\\
& \overline{\zeta_y(t_1) \zeta_y(t_2)} = 2 T_y \delta(t_1 - t_2)  \,.
\end{align}
where an overbar denotes averaging over the different realizations of $\zeta_x(t)$ and $\zeta_y(t)$, while $T_x$ and $T_y$ are the temperatures of the baths.
The trajectories of the BG, which may explore the whole plane $(x,y)$, will be denoted by $\rho_{t} = (x_t,y_t)$ .
\\
We are going to investigate the case in which $T_x$ is vanishingly small, while $T_y$ is finite, which makes the dynamics of the $x$-component 
enslaved by the $y$-component. Several remarks are in order:
\begin{itemize}
\item[\bf (i)] because $|u| < 1$, 
the minimum $O_m =(x_m,y_m)$ of the overall
potential $U(x,y) = U_0(x,y) - F_x x - F_y y$ is located at
\begin{align}
\label{center}
x_m = \frac{F_x - u F_y}{1 - u^2} \,, \quad y_m = \frac{F_y - u F_x}{1 - u^2} \,,
\end{align}
{\em i.e.}\ it shifts away from the origin, when $F_x \neq 0$ or $F_y \neq 0$. 
After a short transient, during which the individual trajectories $x_t$ and $y_t$ drift from their initial positions towards $x_m$ and $y_m$, the trajectory 
$\rho_t(x_t,y_t)$ describes irregular rotations around $O_m$ with non-zero (average) angular velocity $\omega$.  The value of $\omega$ can be inferred 
from the analysis of the curl of the stationary position probability density function, $P_{\infty}(x,x)$ \cite{Lam3}, which is the $t \to \infty$ limit of the 
transient one, $P_{t}(x,y)$. 
\item[\bf (ii)] 
Moreover $|u| < 1$ implies that the solution $P_{t}(x,y)$ of the Fokker-Planck equation has a Gaussian 
$t \to \infty$ limit.  
\item[\bf (iii)]
A more general expression of $U_0$, with arbitrary coefficients multiplying $x^2$ and $y^2$ has been studied {\em e.g.}\
in Refs.\cite{al1,al2,3,sesta}. Qualitatively, the behaviour is essentially the same, although the resulting expressions become fairly more complicated. 
Therefore, for the sake of simplicity we keep Eq.\eqref{pot}. 
\item[\bf (iv)]
The model studied in Refs.\cite{fakhri1,fakhri2,komura} is  equivalent to a biased BG. Indeed, it involves 
 two beads linked by a harmonic spring, which are in contact with two separate heat paths. In addition, the first behaviour is 
 tethered by a harmonic spring to a fixed point at distance $A$ from the origin, while the second is tethered to a point at 
 a larger distance $B$. Thus, the potential takes the form:
\begin{align}
U_b(x,y) = \frac{\kappa'}{2}\left(A - x\right)^2 + \frac{\kappa}{2}\left(y - x\right)^2 + \frac{\kappa'}{2}\left(B - y\right)^2 \,.
\end{align}
Then, suitably rescaling time and temperature, the Langevin equations describing the evolution of the coordinates of the beads
attain the form of equations \eqref{1}, with
\begin{align}
F_x &= \frac{\kappa' A}{\kappa + \kappa'} \,, \quad F_y = \frac{\kappa' B}{\kappa + \kappa'} \,, \quad
u = - \frac{\kappa}{\kappa + \kappa'}  \,.
\end{align} 
As a consequence, the results of Ref.\cite{Lam3}, for the steady-state probability density function, its curl, the angular velocity etc, as well as
our present investigation for $T_x=0$, hold for this model. 
\end{itemize}
Let us introduce the {\em asymmetry function} ${\cal A}_t$, 
\begin{equation}
\label{11}
{\cal A}_t(x,y) = \ln \frac{P_t(x,y)}{P_t(-x,-y)}  
\end{equation}
whose $t \to \infty$ limit has been shown to take the form \cite{Lam3}:
\begin{equation}
{\cal A}_\infty(x,y) = \frac{2 F_x x}{T_{\rm eff}^{(x)}}+ \frac{2  F_y y}{T_{\rm eff}^{(y)}} +  
 u\Bigg( \frac{1}{T _{\rm eff}^{(y)}} -  \frac{1}{T _{\rm eff}^{(x)}} \Bigg)  \left(F_y x - F_x y \right)  \, ,
 \label{FRA}
\end{equation}
where $F_x x$ and $F_y y$ are the works done by ${\bf F}$ along the directions $x$ and $y$, for points starting at $(0,0)$, and
\begin{equation}
\label{temps}
T_{\rm eff}^{(x)} = T_x + \frac{u^2}{4} \frac{\left(T_x - T_y\right)^2}{ T_y } \,,  \qquad
T _{\rm eff}^{(y)}= T_y + \frac{u^2}{4} \frac{\left(T_x - T_y\right)^2}{T_x } \,. 
\end{equation}
are interpreted as {\em effective temperatures}. Evidently, $T_{\rm eff}^{(x)}$ and $T_{\rm eff}^{(y)}$ have no thermodynamic meaning, but they still
suggestively characterize the state of the system, in situations in which thermodynamics does not apply. They can in fact be deduced from the statistics of 
the position distribution.
\\
In the case in which $T_x=0$, one obtains $T^{(x)}_{\rm eff} = u^2 T_y/4$, but the above makes no sense, because $T^{(y)}_{\rm eff}$ is not defined.
Nevertheless, one may either consider the small values of $T_x$, or, given that the effective temperatures concern the steady state, one may investigate
a time dependent situations. In particular, one may introduce a quantity $T^{(y)}_{\rm eff}(t)$ that converges in time to $T^{(y)}_{\rm eff}$ when 
$T_x \ne 0$, and investigate the rate at which it diverges, when $T_x=0$.
\\
This phenomenon deserves some consideration. Indeed, apart from its applicability to systems such as those of Refs.\cite{fakhri1,fakhri2,komura}, it is 
analogous to behaviours reported elsewhere in the literature. For instance, 
analogously to the divergence of $T^{(y)}_{\rm eff}$, Ref.\cite{david}, investigating  the directionality 
of interactions between cellular processes, reported the divergence,  for vanishing noise intensity, of the transfer entropy $T_{W \to V}$, which quantifies the 
information contained in the past of the process $W$ about the future of $V$,  not provided by the  past of $V$. As the phenomena of Ref.\cite{david} 
are rather more complex than ours, one may expect that the divergence of pertinent physical parameters is fairly general. Furthermore, also the dynamics
of populations studied in Ref.\cite{kamenev}, and those of a pair of noisy Kuramoto oscillators living at different temperatures, of Ref.\cite{njp},
exhibit peculiar behaviours in case either of the temperatures is null.

\section{Time-dependence of the effective temperatures}
\label{results}

Without significant lack of generality, let us assume that $x_{t =0} = 0$ and $y_{t=0} = 0$. 
Solving Eqs.\eqref{1} for a given realization of the noise $\zeta_y(t)$ (with $\zeta_x(t) \equiv 0$) we get
\begin{align}
x_t  &= x_m - {\cal R}_+(t) + {\cal R}_-(t) - e^{-t} \int^t_0 d\tau \, \zeta_y(\tau) \, e^{\tau} \, \sinh\left(u (t - \tau)\right) \,, \nonumber\\
y_t  &= y_m - {\cal R}_+(t) - {\cal R}_-(t) + e^{-t} \int^t_0 d\tau \, \zeta_y(\tau) \, e^{\tau} \, \cosh\left(u (t - \tau)\right) \,,
\end{align}
where
\begin{align}
 {\cal R}_+(t) = \frac{F_y + F_x}{2 (1 + u)} e^{-(1 + u) t} \,, \quad 
 {\cal R}_-(t) = \frac{F_y - F_x}{2 (1 - u)} e^{-(1 - u) t} \,.
\end{align}
The corresponding moment-generating function, defined by:
\begin{align}
\Phi(\omega_x,\omega_y) = \overline{\exp\left(i \omega_x x_t + i \omega_y y_t\right)} 
\end{align}
where the overline denotes averaging over of realizations of noise, can be expressed as:
\begin{align}
\label{gen}
&\Phi(\omega_x,\omega_y) = \exp\Big(i \omega_x \left(x_m - {\cal R}_+(t) + {\cal R}_-(t)\right) + i \omega_y \left(y_m - {\cal R}_+(t) - {\cal R}_-(t) \right)      \Big) \nonumber\\
&\times  \overline{\exp\left(i \, e^{-t} \int^t_0 d\tau \, \zeta_y(\tau) \, e^{\tau} \left[\omega_y \cosh\left(u (t - \tau)\right)  - \omega_x   \sinh\left(u (t - \tau)\right)\right]\right)} \nonumber\\
 &= \exp\Big(i \omega_x \left(x_m - {\cal R}_+(t) + {\cal R}_-(t)\right) + i \omega_y \left(y_m - {\cal R}_+(t) - {\cal R}_-(t) \right)      \Big) \nonumber\\
& \times \exp\Big(- T_y \, a_t \, \omega_x^2 - T_y \, b_t \, \omega^2_y - T_y \, c_t \, \omega_x \, \omega_y\Big) \,,
\end{align}
with
\begin{align}
\label{functions}
a_t & = \frac{u^2}{4 (1-u^2)} + \frac{1}{4} e^{- 2 t} - \frac{e^{- 2 t} \left(\cosh\left(2 u t\right) + u \, \sinh\left(2 u t\right)\right)}{4 (1 - u^2)} \,, \nonumber\\
b_t & = \frac{2 - u^2}{4 (1-u^2)} - \frac{1}{4} e^{- 2 t} - \frac{e^{- 2 t} \left(\cosh\left(2 u t\right) + u \, \sinh\left(2 u t\right)\right)}{4 (1 - u^2)} \,, \nonumber\\
c_t & = - \frac{ u}{2 (1-u^2)} + \frac{e^{- 2 t} \left(u \, \cosh\left(2 u t\right) + \sinh\left(2 u t\right)\right)}{2 (1 - u^2)} \,. 
\end{align}
Performing next the inverse Fourier transform of $\Phi$, the time-dependent probability density function is found to be the following Gaussian function:
\begin{align}
\label{dist}
P_t(X,Y) &= \frac{1}{2 \pi  \sqrt{4 a_t b_t - c_t^2} T_y} \nonumber\\
&\times \exp\Bigg( - \frac{b_t \,  \left(X - x_m + {\cal R}_+(t) - {\cal R}_-(t)\right)^2 + a_t \,  \left(Y - y_m + {\cal R}_+(t) + {\cal R}_-(t)\right)^2}{\left(4 a_t b_t - c_t^2\right) T_y} \nonumber\\
&+ \frac{c_t \left(X - x_m + {\cal R}_+(t) - {\cal R}_-(t)\right) \left(Y - y_m + {\cal R}_+(t) + {\cal R}_-(t)\right)}{\left(4 a_t b_t - c_t^2\right) T_y}
\Bigg) \,.
\end{align}
This allows us to attribute a clear physical meaning to the time-dependent parameters defined in \eqref{functions}. In fact, the the low order moments of $P_t(X,Y)$ 
are expressed by:
\begin{align}
\label{1moment}
\overline{x_t} & = x_m  + {\cal R}_{-}(t) - {\cal R}_+(t) \,,\nonumber\\
\overline{y_t} & = y_m - {\cal R}_{-}(t) - {\cal R}_+(t)  \,,
\end{align}
and
\begin{align}
\label{2moment}
\sigma^2_X  = \overline{\left(x_t-\overline{x_t}\right)^2} &= 2 a_t T_y \,, \nonumber\\
\sigma^2_Y = \overline{\left(y_t-\overline{y_t}\right)^2} &= 2 b_t T_y \,,\nonumber\\
\overline{\left(x_t - \overline{x_t}\right) \left(y_t-\overline{y_t}\right)} &= c_t T_y \, .
\end{align}
which means that, multiplying the quantities $a_t$, $b_t$ and $c_t$ of Eqs.\eqref{functions} by $T_y$ -- the non-zero temperature in the system -- 
gives the variances and the cross-correlations of the random processes $x_t$ and $y_t$. Moreover, the time-dependent asymmetry function reads:
\begin{align}
{\cal A}_t = \ln \frac{P_t(X,Y)}{P_t(-X,-Y)} &= \frac{2 F_x X}{T^{(x)}_{\rm eff}(t)} +  \frac{2 F_y Y}{T^{(y)}_{\rm eff}(t)} + \nonumber\\
&+ \left(\frac{1}{T^{(y)}_{\rm eff}(t)} - \frac{1}{T^{(x)}_{\rm eff}(t)}\right) \left(f_1 F_y X - f_2 F_x Y\right) \,,
\label{ATT}
\end{align}
where we have introduced the time dependent effective temperatures
\begin{align}
\label{temperatures}
T^{(x)}_{\rm eff}(t) &= \dfrac{\left(1 - u^2\right) \left(4 a_t b_t - c_t^2\right) T_y}{2 b_t + u c_t - \dfrac{\left(1+u\right)}{2} \left(2 b_t + c_t\right) e^{-(1-u) t} + \dfrac{\left(1 - u\right)}{2} \left(c_t - 2 b_t\right) e^{- (1+ u) t}}  \,,\nonumber\\
T^{(y)}_{\rm eff}(t) &= \dfrac{\left(1 - u^2\right) \left(4 a_t b_t - c_t^2\right) T_y}{2 a_t + u c_t - \dfrac{\left(1+u\right)}{2} \left(2 a_t + c_t\right) e^{-(1-u) t} + \dfrac{\left(1 - u\right)}{2} \left(c_t - 2 a_t\right) e^{- (1+ u) t}} \,,
\end{align}
and the time dependent parameters
\begin{align}
f_1 = \dfrac{2 b_t + c_t}{b_t - a_t} - \dfrac{2 (1 - u)  \left(1 - e^{-(1+ u) t}\right) b_t}{\left(b_t - a_t\right) \left(1 - \dfrac{\left(1+u\right)}{2} e^{-(1-u) t} - \dfrac{\left(1 - u\right)}{2} e^{-(1+u)t} \right)} \,, \nonumber\\
f_2 = - \dfrac{2 a_t + c_t}{b_t - a_t} + \dfrac{2 (1 - u)  \left(1 - e^{-(1+ u) t}\right) a_t}{\left(b_t - a_t\right) \left(1 - \dfrac{\left(1+u\right)}{2} e^{-(1-u) t} - \dfrac{\left(1 - u\right)}{2} e^{-(1+u)t} \right)} \,.
\end{align}
The expressions in Eqs.\eqref{temperatures}, which are valid at all times $t$, constitute the main result of the present paper. 
\\
We also realize that the mean values of $x_t$ and $y_t$, which initially vanish, approach their asymptotic values $x_m$ and $y_m$ exponentially fast, see 
Eq.\eqref{1moment}). In turn, the variances of $x_t$ and $y_t$, which also initially vanish, grow monotonically with $t$, 
and eventually saturate at constant values. In the large-$t$ limit, they obey: 
\begin{align}
\sigma^2_X  &=  \frac{u^2 T_y}{2 \left(1 - u^2\right)} - \frac{T_y}{4(1 - |u|)} e^{-2 (1 - |u|) t}  + o\left(e^{-2 (1 - |u|) t}\right)  \,, \nonumber\\
\sigma^2_Y  &=   \frac{(2 - u^2) T_y}{2 \left(1 - u^2\right)} - \frac{T_y}{4(1 - |u|)} e^{-2 (1 - |u|) t}  + o\left(e^{-2 (1 - |u|) t}\right) \,.
\end{align} 
Note that the variance of the driving component, $y_t$, is always bigger than the variance of the passive one, which is understandable. 
In turn, also the cross-correlation between the two components evolves towards a constant value, {\em i.e.}\ they do not decouple in the
steady-state!  Asymptotically in $t$, it goes like:
\begin{align}
\overline{\left(x_t - \overline{x_t}\right) \left(y_t-\overline{y_t}\right)} = - \frac{u T_y}{2 (1-u^2)} + \frac{{\rm sign}(u) T_y}{4 (1 - |u|)} e^{-2 (1 - |u|) t}  + o\left(e^{-2 (1 - |u|) t}\right) \,.
\end{align}
the cross correlation function is finite at all times $t$, for $|u| < 1$, $T_x = 0$ and $T_y > 0$. Its limit value is negative for $u > 0$, and positive, otherwise.
\\
Concerning the time-evolution of the effective temperatures, Eqs.\eqref{temperatures}, we observe that $T^{(x)}_{\rm eff}(t)$ monotonically increases 
in time: initially, it correctly vanishes, since the thermodynamic temperature of the passive component is, $T_x=0$, and then it saturates at a positive value 
as $t \to \infty$. Its asymptotic behaviour in time goes like:
\begin{align}
T^{(x)}_{\rm eff}(t)  = \frac{u^2}{4} T_y + \frac{u^2 (2 + |u|) T_y}{16} e^{-(1 - |u|) t} + o\left( e^{-(1 - |u|) t}\right) \,.
\end{align}
It is intriguing that the limiting value of $T^{(x)}_{\rm eff}(t)$ is just  $1/4$, or less, of the thermodynamic temperature of the active component,
$T_y$, its maximum being achieved as $u$ approaches $+1$ or $-1$. At the same time, the effective temperature of the driving component, also
monotonically increases with time. Correctly, it initially equals the thermodynamic temperature, $T^{(y)}_{\rm eff}(t) = T_y$, and then it grows 
without bounds, diverging as the system tends to a steady-state, consistently with \eqref{temps},cf.\ \cite{Lam3}. Its asymptotic behaviour obeys:
\begin{align}
T^{(y)}_{\rm eff}(t) = |u| T_y e^{(1 - |u|) t} - \frac{T_y}{|u|} e^{-(1 - |u|)t } + o\left(e^{-(1 - |u|)t }\right) \,,
\end{align}
which shows that $T^{(y)}_{\rm eff}(t)$ grows exponentially fast in time.

\section{Conclusions}
\label{conc}
In this paper we have investigated a notion of effective temperatures for nonequilibrium BG, and their time evolution, considering in particular the case 
in which one variable is coupled to a zero-temperature bath. The usefulness of effective temperatures may be questioned, as pointed out {\em e.g.}\
In Ref.\cite{VBPV}. The point is that apparent violations of the Fluctuation Dissipation Relation may arise, when in reality one has not considered the proper 
degrees of freedom (maybe because experimentally inaccessible, but present). Consequently some improper correlation function may be compared with the 
response function of interest, leading to the conclusion that the FDR is violated, when it is not. In such cases, then, restoring the FDR by means of an
effective temperature may be scarcely revealing; better investigating the presence of extra degrees of freedom, and identifying the proper correlation functions 
\cite{VBPV}. 
\\
However, this approach is not always viable. Moreover, we adopted a different point of view. Our asymmetry function ${\cal A}_t$, which only superficially looks
like the argument of the Fluctuation Relations, is computable and in principle experimentally accessible, in terms of the time dependent or the stationary
distribution of positions $(x,y)$. The result can be compared with the right hand side of Eq.\eqref{ATT} or Eq.\eqref{FRA}, respectively. The parameters
$T^{(x)}_{\rm eff}$ and $T^{(y)}_{\rm eff}$ there introduced, are called effective temperatures because, apart from the Boltzmann constant $k_{_B}$
assumed to be 1 in this paper, have the dimension of temperature and reduce to the single temperature, $T_x=T_y$, when in equilibrium. Moreover, they
meaningfully characterize the state of the BG. For instance, suppose that only one of the temperatures of the baths is known. Then, Eqs.\eqref{ATT} and 
Eq.\eqref{FRA} can be used to infer the value of the other one, observing the value or the behaviour of ${\cal A}_t$. In particular, the case in which
one of the baths is at vanishing or very small temperature would be evidenced by an exponential time dependence.
Analogously, knowledge of the temperatures, and of a single component of the driving force, allows the calculation of the other component. 
Furthermore, there is also the case in which ${\cal A}_t$, hence the statistic of positions, is not known, or only partially known ({\em e.g.}\ $P_t$ is known
only for positive $X$), while the right hand sides of Eq.\eqref{ATT} or of Eq.\eqref{FRA} are known. Indeed, these right hand sides only require knowledge 
of the BG parameters. Again, Eqs.\eqref{ATT} and \eqref{FRA} can be used to obtain additional information on the position statistic.
\\
Because BG are realized in practice, these facts can be used to understand their behaviour and to properly tune them for practical applications.


\section{References}

\begin{thebibliography}{99}

\bibitem{Kubo91} R. \ Kubo, M. \ Toda, N. \ Hashitsume, {\it Statistical Physics II. Nonequilibrium Statistical Mechanics}, Springer-Verlag, Berlin (1991)

\bibitem{EvMo} G.P.\ Morriss, D.J.\ Evans, {\it Statistical Mechanics of Non--equilibrium Liquids},
Cambridge University Press (2008)

\bibitem{Gall14} G.\ Gallavotti, {\it Nonequilibrium and Irreversibility}, Springer-Verlag, Berlin (2014)

\bibitem{Ciliberto} S.\ Ciliberto, 
{\it Experiments in Stochastic Thermodynamics: Short History and Perspectives}, Phys.\ Rev.\ X {\bf 7} 021051 (2017)

\bibitem{ECM2} D.J.\ Evans, E.G.D.\ Cohen and G.P.\ Morriss, {\it Probability of second law violations in 
shearing steady flows}, Phys.\ Rev.\ Lett.\ {\bf 71} 2401 (1993)

\bibitem{GGBol} G.\ Gallavotti, {\it Ergodicity, ensembles and beyond}, J.\ Stat.\ Phys.\ {\bf 78}
1571 (1995)

\bibitem{Jarz} C.\ Jarzynski, {\it Nonequilibrium equality for free energy differences}, 
Phys.\ Rev.\ Lett.\ {\bf 78} 2690 (1997)

\bibitem{ES2002} D.J.\ Evans, D.J.\ Searles, {\it The fluctuation theorem}, Adv.\ Phys.\ {\bf 52} 1529 (2002)

\bibitem{J2004}
C.\ Jarzynski, {\it Nonequilibrium work theorem for a system strongly coupled to a thermal environment},
J.\ Stat.\ Mech.\ P09005 (2004)

\bibitem{Bettolo} B.M.U.\ Marini, A.\ Puglisi, L.\ Rondoni, A.\ Vulpiani, {\it Fluctuation–dissipation: response
theory in statistical physics}, Phys.\ Rep.\ {\bf 461} 111 (2008)

\bibitem{Kirkp} T.R.\ Kirkpatrick, E.G.D.\ Cohen, J.R.\ Dorfman, {Light scattering by a fluid in a nonequilibrium steady state. II. Large gradients},
Phys.\ Rev.\ A {\bf 26} 995 (1982)

\bibitem{Dorfm} J.R.\ Dorfman, J.R.\ Kirkpatrick, J.V.\ Sengers, {\it Generic Long-Range Correlations in Molecular Fluids}, 
Ann.\ Rev.\ Phys.\ Chem.\ {\bf 45} 213 (1994)

\bibitem{Cugli97} L.F.\ Cugliandolo, J.\ Kurchan, L.\ Peliti,  {\it Energy flow, partial equilibration, and effective temperatures in systems with slow dynamics},
Phys.\ Rev.\ E {\bf 55} 3898 (1997)

\bibitem{Ortiz} J.M.\ Ortiz de Z\'arate, J.V.\ Sengers, {\it Hydrodynamic Fluctuations in Fluids and Fluid Mixtures},
Elsevier, Amsterdam (2006)

\bibitem{ContiD} L.\ Conti, P.\ De Gregorio, G.\ Karapetyan, C.\ Lazzaro, M.\ Pegoraro, M.\ Bonaldi, L.\ Rondoni,
{\it Effects of breaking vibrational energy equipartition on measurements of temperature in macroscopic oscillators subject to heat flux},
J.\ Stat.\ Mech.\ P12003 (2013)

\bibitem{peter} P. H\"anggi and F. Marchesoni, {\em Artificial Brownian motors: Controlling transport on the nanoscale},  Rev. Mod. Phys. {\bf 81}, 387 (2009).

\bibitem{prost} M. C. Marchetti, J. F. Joanny, S. Ramaswamy, T. B. Liverpool, J. Prost, Madan Rao, and R. Aditi Simha, {\em Hydrodynamics of soft active matter},
Rev. Mod. Phys. {\bf 85}, 1143 (2013).

\bibitem{Scholz} C.\ Scholz, S.\ Jahanshahi, A.\ Ldov, H. L\"owen, {\it Inertial delay of self-propelled particles}, Nat.\ Commun.\ {\bf 9}, 5156 (2018)


\bibitem{nos1} D. Helbing, {\em Traffic and related self-driven many-particle systems}, Rev. Mod. Phys. {\bf 73},  1067 (2001).

\bibitem{nos2} A. V. Ivlev, J. Bartnick, M. Heinen, C.-R. Du, V. Nosenko, and
H. L\"owen, {\em Statistical Mechanics where Newton's Third Law is Broken}, Phys. Rev. X {\bf 5}, 011035 (2015).

\bibitem{carl1} C. Mejia-Monasterio and G. Oshanin, {\em Bias-and bath-mediated pairing of particles driven through a quiescent medium}, Soft Matter {\bf 7}, 993 (2011).

\bibitem{vinc} A. Poncet, O. B\'enichou, V. D\'emery, and G. Oshanin, {\em Universal long ranged correlations in driven binary mixtures}, Phys. Rev. Lett.  {\bf 118}, 118002 (2017).

\bibitem{carl2} O. A. Vasilyev, O. B\'enichou, C. Mejia-Monasterio, E. R. Weeks, and G. Oshanin, {\em Cooperative behaviour of biased probes in crowded interacting systems}, Soft Matter {\bf 13}, 7617 (2017).

\bibitem{oli} A. Poncet, O. B\'enichou, V. D\'emery, and G. Oshanin, {\em Bath-mediated interactions between driven tracers in dense single files},  Phys. Rev. Research {\bf 1}, 033089 (2019).

\bibitem{evans} D. J. Evans and D. J. Searles, {\em The fluctuation theorem},
Adv. Phys. 51, 1529 (2002).

\bibitem{Lam1} L. Rondoni and C. Mej\'{i}a-Monasterio, {\em Fluctuations in non-equilibrium
statistical mechanics: models, mathematical theory,
physical mechanisms}, Nonlinearity {\bf 20}, R1 (2007).

\bibitem{Lam2} U. M. B. Marconi, A. Puglisi, L. Rondoni, and A. Vulpiani,
{\em Fluctuation-dissipation: Response theory in statistical physics},
Phys. Rep. {\bf 461}, 111 (2008).

\bibitem{Seki} K. Sekimoto, {\em Stochastic Energetics}, (Springer-Verlag, Berlin, Heidelberg, 2010).

\bibitem{Udo} U. Seifert, {\em Stochastic thermodynamics, fluctuation theorems, and molecular machines}, Rep. Progr. Phys. {\bf 75}, 126001 (2012).



\bibitem{pug} A. Puglisi, A. Sarracino, and A. Vulpiani, {\em Temperature in and
out of equilibrium: A review of concepts, tools and attempts},
Phys. Rep. {\bf 709-710}, 1 (2017).

\bibitem{ol1} O. B\'enichou, P. Illien, G. Oshanin, A. Sarracino, and R Voituriez, {\em Nonlinear response and emerging nonequilibrium microstructures for biased diffusion in confined crowded environments}, Phys. Rev. E {\ bf 93}, 032128 (2016).

\bibitem{ol0} P. Illien, O B\'enichou, G. Oshanin, A. Sarracino, and R. Voituriez, {\em Nonequilibrium fluctuations and enhanced diffusion of a driven particle in a dense environment}, Phys. Rev. Lett. {\bf 120}, 200606 (2018).



\bibitem{ol} O B\'enichou, P. Illien, G. Oshanin, A. Sarracino, and R. Voituriez, {\em Tracer diffusion in crowded narrow channels}, J. Phys.: Condens. Matter {\bf 30}, 443001 (2018).

\bibitem{neri} I. Neri, E. Rold\'an,  
and F. J\"ulicher, {\em Statistics of Infima and Stopping Times of Entropy Production and Applications to Active Molecular Processes}, Phys. Rev. X {\bf 7}, 011019 (2017).


\bibitem{neri2} I. Neri, E. Rold\'an, S. Pigolotti, 
and F. J\"ulicher, {\em Integral fluctuation relations for
entropy production at stopping times}, J. Stat. Mech. (2019) 104006.

\bibitem{pel} R. Exartier and L. Peliti, {\em A simple system with two temperatures},
Phys. Lett. A {\bf 261}, 94 (1999).


\bibitem{1} R. Filliger and P. Reimann, {\em Brownian Gyrator: A Minimal Heat Engine on the Nanoscale}, 
Phys. Rev. Lett. {\bf 99}, 230602 (2007).



\bibitem{al1} S. Ciliberto, A. Imparato, A. Naert and M. Tanase, {\em Heat Flux and Entropy Produced by Thermal Fluctuations}, Phys. Rev. Lett. {\bf 110}, 180601 (2013).

\bibitem{al2} S. Ciliberto, A. Imparato, A. Naert and M. Tanase, {\em Statistical properties of the energy exchanged between two heat baths coupled by thermal fluctuations}, J. Stat. Mech. {\bf 2013}, P12014

\bibitem{crisanti} A. Crisanti, A. Puglisi and D. Villamaina, {\em Nonequilibrium and information: The role of cross correlations}, Phys. Rev. E {\bf 85}, 061127 (2012).

\bibitem{2} V. Dotsenko, A. Maciolek, O. Vasilyev, and G. Oshanin, {\em Two-temperature Langevin dynamics in a parabolic potential},
Phys. Rev. E {\bf 87}, 062130 (2013). 

\bibitem{12} A. Yu. Grosberg and J.-F. Joanny, {\em Nonequilibrium statistical mechanics of mixtures of particles in contact with different thermostats}, 
Phys. Rev. E 92 032118 (2015).

\bibitem{11}  A. Argun, J. Soni, L. Dabelow, S. Bo, G. Pesce, R. Eichhorn,
and G. Volpe, {\em Experimental realization of a minimal microscopic
heat engine}, Phys. Rev. E 96, 052106 (2017).

\bibitem{david} S. Lahiri, P. Nghe, S. J. Tans, M. L. Rosinberg, and D. Lacoste, {\em Information-theoretic analysis of the directional influence between cellular processes}, PLoS ONE {\bf 12}, e0187431 (2017).

\bibitem{3} V. Mancois, B. Marcos, P. Viot, and D. Wilkowski, {\em Two-temperature Brownian dynamics of a particle in a confining potential}, Phys. Rev. E {\bf 97}, 052121 (2018).


\bibitem{Lam3} S. Cerasoli, V. Dotsenko, G. Oshanin, and L. Rondoni, {\em Asymmetry relations and effective temperatures for biased Brownian gyrators}, Phys. Rev. E {\bf 98}, 042149 (2018).


\bibitem{fakhri1} C. Battle, C. P. Broedersz, N. Fakhri, V. F. Geyer,
J. Howard, C. F. Schmidt, and F. C. MacKintosh, {\em Broken detailed balance at
mesoscopic scales in active
biological systems}, Science {\bf 352}, 604 (2016).

\bibitem{fakhri2} J. Li, J. M. Horowitz, T. R. Gingrich and  N. Fakhri, {\em Quantifying dissipation using fluctuating currents}, Nat. Comm. {\bf 10}, 1666 (2019).   

\bibitem{komura} I. Sou, Y. Hosaka, K. Yasuda, and S. Komura, {\em Non-equilibrium probability flux of a thermally driven micromachine}, Phys. Rev. E {\bf 100}, 022607 (2019).

\bibitem{sesta} A. Baldassarri, A. Puglisi, and L. Sesta, {\em Engineered Swift Equilibration of a Brownian Gyrator}, Phys. Rev. E {\bf 102}, 030105(R)  (2020).

\bibitem{nas} E. dos S. Nascimento and  W. A. M. Morgado, {\em Memory effects on two-dimensional overdamped Brownian dynamics}, J. Phys. A: Math. Theor. {\bf 53}, 065001 (2020);
{\em Stationary properties of a Brownian gyrator with non-Markovian baths};  arXiv:2002.07254v2

\bibitem{kamenev} M. Parker, A. Kamenev, and B. Meerson, {\em Noise-Induced Stabilization in Population Dynamics}, Phys. Rev. Lett. {\bf 107},180603 (2011).

\bibitem{njp} V. Dotsenko, A. Maciolek, G. Oshanin, O. Vasilyev, and S. Dietrich, {\em Current-mediated synchronization of a pair of beating non-identical
flagella}, New J. Phys. {\bf 21},   033036 (2019).

\bibitem{VBPV} D.\ Villamaina, A.\ Baldassarri, A.\ Puglisi, A.\ Vulpiani, {\it The fluctuation-dissipation relation:
how does one compare correlation functions and responses?}, J.\ Stat.\ Mech.\  P07024 (2009) 






\end{thebibliography}
\end{document}